# Is GGAG:Ce@SiO$_2$-RB composite a prospective material for X-ray induced photodynamic therapy?

Iveta Terezie Hošnová[a], Kristýna Havlinová[a]*, Jan Bárta[a,b], Karolína Mocová[a], Xenie Lytvynenko[a], Lenka Prouzová Procházková[a,b], Vojtěch Kazda[a], František Hájek[b], Viliam Múčka[a], Václav Čuba[a] a     Faculty of Nuclear Sciences and Physical Engineering, Czech Technical University in Prague, Břehová 7, 115 19 Prague 1, Czech Republic

b     Institute of Physics, Czech Academy of Sciences, Cukrovarnická 10, 162 00 Prague 6, Czech Republic

**Abstract**

Nanocomposite material (GGAG:Ce$^{3+}$@SiO$_2$-RB) for potential use in X-ray induced photodynamic therapy (X-PDT) was developed, thoroughly characterized, and evaluated. It consists of a scintillating Gd$_3$(Ga$_{1-x}$Al$_x$)$_5$O$_{12}$:Ce$^{3+}$ core encapsulated in silica layer and functionalized with the photosensitizer Rose Bengal (RB). Radioluminescence measurements confirmed the energy transfer from the scintillating core to Rose Bengal. Dark toxicity and radiosensitisation effects were evaluated using *Saccharomyces cerevisiae* as a model organism. The nanocomposite showed minimal dark toxicity at concentrations of up to 10 mg/mL. However, X-ray irradiation experiments did not demonstrate significant singlet oxygen production compared to the controls. Although the nanocomposite design shows potential, further optimization is needed to achieve an effective X-PDT performance.

## 1. Introduction

Radiodynamic therapy (RDT) is an emerging anticancer treatment that combines the principles of radiotherapy and photodynamic therapy. It involves the use of ionizing radiation to activate photosensitizer drugs, generating reactive oxygen species (ROS) that cause damage to tumour cells (Clement et al., 2021). X-ray induced photodynamic therapy (X-PDT) is a variant of RDT that typically involves the use of composite nanoparticles containing both scintillators and photosensitizers. These scintillators convert X-ray energy into visible light, which activates photosensitizers to generate ROS (Wang et al., 2016; Wang et al., 2022). X-PDT has shown potential for treating various types of cancer, particularly deep-seated tumours that are difficult to treat with conventional photodynamic therapy or radioresistant tumours, including colorectal cancer, pancreatic cancer, glioblastoma, or lung tumours (Daouk et al., 2021; Sang et al., 2022; Yang et al., 2019). Previous studies have explored a range of inorganic scintillators for X-PDT, including $SrAl_2O_4$:Eu (Chen et al., 2015), $LaF_3$:Ce (Bulin et al., 2020), $LaF_3$:Tb (Liu et al., 2008), $LiLuF_4$:Ce (Wang et al., 2022), $Lu_3Al_5O_{12}$:Pr (Popovich et al., 2018), $Tb_2O_3$ (Bulin et al., 2013), $NaLuF_4$:Dy,Gd (Nsubuga et al., 2021), $Gd_3Al_5O_{12}$:Ce (Jain et al., 2018a), and many others.

Multicomponent garnets, such as $Gd_3Ga_3Al_2O_{12}$ (GGAG), have attracted significant attention in both single crystal and powder / nanoparticle forms due to their versatile properties and applications. Garnets are, like other cubic materials, optically isotropic, which allows the fabrication of high-quality optical ceramics and single crystals (Wu et al., 2015; Yanagida et al., 2013). The garnet structure provides a stable host for various dopants, such as $Ce^{3+}$ or other lanthanide ions, which can enhance its optical and scintillation properties (Bárta et al., 2018).

In single crystal form, GGAG:Ce exhibits excellent scintillation characteristics, including a high light yield of 45 000±2 500 photons/MeV under 662 keV γ-ray excitation and an energy resolution of 7.6% for 662 keV γ-rays (Wu et al., 2015). These properties make GGAG single crystals promising candidates for applications in radiation detection and imaging. Nanosized GGAG can be prepared from aqueous solutions using an indirect photochemical synthesis method (Bárta et al., 2018). This approach involves the following two steps: preparation of solid precursors from aqueous solutions using photochemical synthesis and calcination of the prepared precursors under various conditions to produce the final GGAG nanoparticles. The properties of the GGAG nanoparticles can be tailored mostly by the Ga content, which has a strong impact on the lattice parameter and position of the $Ce^{3+}$ 5d–4f emission. The addition of $Mg^{2+}$ ions to the precursor solutions and/or calcination in air can induce the formation of $Ce^{4+}$ ions as defects, which compete with electron traps and cause an acceleration of luminescence decay at the cost of decreased $Ce^{3+}$ emission intensity (Bárta et al., 2018). This suggests that these factors can be used to fine-tune the optical properties of GGAG nanoparticles for specific applications. One of the potential applications is X-PDT utilising photosensitizers with suitable properties, such as Rose Bengal.

Rose Bengal (RB) is a widely studied photosensitizer known for its ability to generate singlet oxygen, making it invaluable in photodynamic therapy (PDT) applications for treating cancer and bacterial infections (Dhaini et al., 2022). When exposed to light, RB produces singlet oxygen and other reactive oxygen species that can effectively kill tumour cells and bacteria. While RB is generally considered a Type II photosensitizer (singlet oxygen mechanism), its mechanism of action can vary depending on the experimental conditions. For instance, in the Rose Bengal-sensitized photooxidations of tryptophan and its derivatives, the reaction proceeds exclusively via the Type II mechanism (Inoue et al., 1982). However, when RB is loaded into mesoporous silica-coated nanoparticles and exposed to

X-rays, it may produce superoxide and hydroxyl radicals instead of singlet oxygen, highlighting the importance of using multiple assays to determine the specific reactive oxygen species generated in different systems (Nsubuga et al., 2021). Rose Bengal is currently used in ophthalmology for diagnosing dry eyes and has completed phase II clinical trials as an intralesional injection for the PDT treatment of melanoma (Chen et al., 2018). However, its clinical development for PDT has been limited by insufficient lipophilicity and tumour accumulation (Chang et al., 2008). Ongoing research has focused on developing improved formulations and delivery systems to enhance its efficacy and overcome its pharmacological limitations, potentially leading to more widespread clinical use in the future (Chang et al., 2008; Chen et al., 2018; Dhaini et al., 2022).

Singlet oxygen detection is crucial for evaluating the efficacy of X-PDT. To detect and quantify singlet oxygen production, chemical probes are commonly employed, such as aminophenyl fluorescein (APF) (Price et al., 2009; Takahashi & Misawa, 2009), 9,10-anthracenediyl-bis(methylene)dimalonic acid (ABDA) (Entradas et al., 2020; Felip-León et al., 2019; Nsubuga et al., 2021), 1,3-diphenylisobenzofuran (DPBF) (Entradas et al., 2020; Jain et al., 2018b; Nsubuga, et al., 2021; Wozniak et al., 1991; Żamojć et al., 2017; Zhang & Li, 2011), and Singlet Oxygen Sensor Green (SOSG) (Garcia-Diaz et al., 2016; Kim et al., 2013; Lin et al., 2010, 2013; Liu et al., 2019). These probes react more or less specifically with singlet oxygen, resulting in measurable changes in their fluorescence or absorption spectra. For instance, APF and SOSG exhibit increased fluorescence upon reaction with singlet oxygen, while ABDA and DPBF show a decrease in absorption. Furthermore, DPBF exhibits a reduction in fluorescence intensity following the reaction with singlet oxygen (Wu et al., 2011). By monitoring these spectral changes during the photodynamic process, the singlet oxygen generation efficiency can be assessed. Unfortunately, during X-PDT processes, other reactive oxygen species are produced, for instance, by water radiolysis. Consequently, X-ray irradiation increases the demand for the specificity of chemical probes and the methodology of their usage.

In this work we design a nanocomposite system labelled GGAG:Ce@SiO$_2$-RB as a candidate for X-PDT application. This multicomponent composite contains a scintillating core of Gd$_3$Ga$_{5-5x}$Al$_{5x}$O$_{12}$ garnet doped with cerium (GGAG:Ce). The Ce$^{3+}$ dopant is chosen so that the radioluminescence emission spectrum of the core overlaps with the optical absorption spectrum of the photosensitiser used. To ensure the chemical reactivity of the chemically almost inert core toward photosensitiser or other compounds, it is encapsulated in a silica layer (@SiO$_2$). The photosensitiser Rose Bengal (RB) is chosen for functionalization and is grafted onto the silica layer. This composite nanomaterial has good potential for singlet oxygen production by X-rays, owing to the design methodology successfully employed in the literature (Jain et al., 2018b; Popovich et al., 2016, 2018). The subsequent phase for efficient X-PDT application involves the precise targeting of cancer cells within the human body. To achieve active tumour targeting, the silica shell of the nanoparticle was further functionalized with folic acid (FA), a targeting agent for the folate receptor, which is frequently overexpressed on cancer cells and is known to mediate efficient receptor-driven uptake (Rashidi et al., 2016, Zhang et al., 2002). The control material (SiO$_2$-RB) consisting of nanoparticle SiO$_2$ and photosensitiser RB was synthesised to distinguish the effect of radiosensitisation and photodynamic effect under X-rays. The prepared materials were characterised by various methods, tested for singlet oxygen production, and evaluated for dark toxicity on *Saccharomyces cerevisiae* cell cultures.

## 2. Materials and methods

### 2.1. Nanocomposite synthesis

#### 2.1.1. GGAG:Ce core synthesis

The synthesis of Gd$_3$Ga$_{2.5}$Al$_{2.5}$O$_{12}$:Ce was based on a photo-induced method described in previous studies (Bárta et al., 2012, 2018, 2019). An initial solution was prepared in deionised water in a 2 L volumetric flask, containing 3·10$^{-3}$ mol/L gadolinium nitrate hexahydrate (Gd(NO$_3$)$_3$·6H$_2$O), 2.5·10$^{-3}$ mol/L gallium nitrate nonahydrate (Ga(NO$_3$)$_3$·9H$_2$O), 2.5·10$^{-3}$ mol/L aluminum nitrate nonahydrate (Al(NO$_3$)$_3$·9H$_2$O), and 0.1 mol/L ammonium formate (HCOONH$_4$). The solutions were doped with Ce by adding 3·10$^{-5}$ mol/L cerium nitrate hexahydrate (Ce(NO$_3$)$_3$·6H$_2$O), which corresponds to 1 mol. % of Ce$^{3+}$ relative to rare-earth ions. All chemicals were supplied by Merck (Sigma Aldrich), had a minimum purity of 99.995% (trace metal basis), and were used as received without any treatment. The aqueous solution was irradiated using four low-pressure mercury lamps for approximately 4 h with continuous stirring. The power input of the lamps was 25 W per lamp, and the wavelength of the emitted light was mainly 253.7 nm. A very fine white gelatinous precipitate was formed during irradiation, which was filtered by micro-filtration (Millipore HAWP filter, 0.45 μm) and dried at 40 °C for 24 h in air. The solid product was then calcined at 1200 °C in air for 2 h. For several experiments, Gd$_3$Ga$_2$Al$_3$O$_{12}$:1% Ce nanoparticles synthesized in the same process (with Ga$^{3+}$ and Al$^{3+}$ concentrations changed to 2·10$^{-3}$ mol/L Ga(NO$_3$)$_3$·9H$_2$O and 3·10$^{-3}$ mol/L Al(NO$_3$)$_3$·9H$_2$O, respectively) were used as well due to their increased luminescence intensity. The prepared GGAG:Ce powder was ground in 0.1 mol/L citric acid using a Fritsch pulverisette ball mill for 4 h (4× 1 h grinding with 1 h pauses) in order to break agglomerates formed during calcination. Subsequently, the product was centrifuged out of the dispersion, washed with fresh 0.1 mol/L citric acid, and dried at 40 °C for a minimum of 24 hours in air.

#### 2.1.2. Surface modification by amorphous silica layer

The prepared nanoparticles (NPs) were coated with a silica layer using a modified sol-gel process described by Liu et al. (1998) and modified by Popovich et al. (2018). 300 mg of ground GGAG:Ce powder was suspended in 100 mL of absolute ethanol (≥ 99.8%, p.a., PENTA) and ultrasonicated for

1 h. The suspension was placed on a magnetic stirrer, and 35 μL of tetraethoxysilane (TEOS; ≥ 99%, Merck) was added. Subsequently, 10.3 mL of 25-27% ammonium hydroxide solution (NH$_4$OH) was slowly added dropwise and the mixture was left overnight under stirring to maximise hydrolysis of TEOS and formation of a silica layer. For further functionalization, 249 μL of (3-aminopropyl)triethoxysilane (APTES; ≥98.0%, Merck) was added and stirred overnight at room temperature. The silica-coated nanoparticles or coated nanoparticles modified with APTES were 3 times rinsed with water and centrifuged. Finally, they were rinsed with ethanol and left to dry at 40 °C in air for 24 h.

*2.1.3. Functionalization process with photosensitizer Rose Bengal*

The functionalization process was inspired by Cantelli et al. (2020). In this process, the photosensitizer was first activated with Sulfo-NHS (N-hydroxysulfosuccinimide sodium salt) and EDC (N-(3-dimethylaminopropyl)-N'-ethylcarbodiimide) using EDC/NHS crosslinking of carboxylate with primary amine. 300 mg of Rose Bengal was dissolved in 30 mL of DMSO (p.a., Penta) to obtain a concentration of ~ 10 mmol/L. The reaction flask was placed on a magnetic stirrer, and 98 mg of Sulfo-NHS (≥98.0%, Merck) and 146 μL of EDC (≥97.0%, Merck) were added. The mixture was left to react overnight under stirring at room temperature. Subsequently, 300 mg of coated NPs modified with APTES was dispersed in 50 mL of carbonate buffer and ultrasonicated for 15 min. This suspension was then added to the activated Rose Bengal solution. The mixture was left to react overnight with stirring at room temperature. Functionalized GGAG:Ce@SiO$_2$-RB NPs were washed twice with carbonate buffer, once with water, and finally dispersed in ethanol and dried at 40 °C in air. The same functionalization process was used for SiO$_2$-RB synthesis.

*2.1.4. Functionalization process with folic acid*

Folic acid was chosen as a targeting molecule to exploit the frequent overexpression of folate receptors on various cancer cell types and it was conjugated with NPs using procedure reported by Zhang et al., (2002). For functionalization, three stock solutions were prepared: 10 mmol/L folic acid (≥97.0%, Merck (Sigma Aldrich)) in DMSO (p.a., Penta), 15 mmol/L N-hydroxysuccinimide (NHS; ≥98.0%, Merck (Sigma Aldrich)) in water and 75 mmol/L EDC (≥97.0%, Merck (Sigma Aldrich)) in water. The reaction mixture consisted of 14 mL of folic acid solution, 21 mL of NHS solution, and 21 mL of EDC solution. 300 μL of triethylamine TEA (p.a., Penta) and 200 μg of NPs were added and the mixture was left to react overnight with stirring at 37 °C. Functionalized NPs were washed, centrifuged, and dried at 40 °C in air.

2.2. Characterisation of nanocomposites

X-ray powder diffraction (XRPD) of the prepared materials was measured using Rigaku MiniFlex 600 diffractometer equipped with Cu X-ray tube (Ni-filtered Cu-K$_{\alpha1,2}$ radiation; $\lambda_{av}$ = 0.154184 nm) operated at 40 kV and 15 mA and NaI:Tl scintillation detector. The diffraction patterns were recorded in continuous mode between 10° and 80° 2θ with a collection speed of 2°/min and a data collection interval of 0.02°. The measured patterns were evaluated in the PDXL2 program and compared to the relevant records in the ICDD PDF-2 database of powder diffraction data, version 2013.

A Malvern Zetasizer Ultra Red instrument was employed to conduct dynamic light scattering analyses at ambient temperature. The ethanol-suspended samples were examined using a 173° backscattering configuration. Five replicate measurements were performed for each specimen, with a 120-second equilibration period interspersed between successive analyses. To enhance the accuracy of the results, a fluorescence filter was integrated into the experimental setup. Using a comparable setup, we assessed colloidal stability with three replicates per measurement, recording measurements every hour, for a total time of 90 hours.

The samples were imaged using a high-resolution Tescan MAIA3 scanning electron microscope equipped with detectors for the detection of secondary electrons (SE) and back-scattered electrons (BSE). The SE images were used mainly to assess the particle size and overall morphology, while the BSE images reveal mostly the differences in $Z_{eff}$, particles with heavy elements appearing bright in BSE.

Absorption spectra of Rose Bengal, ABDA probe, and Fricke solution were measured using a dual-beam UV-VIS spectrophotometer Varian Cary 100 with an optical length of 1 cm.

Radioluminescence (RL) spectra were acquired at room temperature using a custom made 5000M spectrofluorometer (Horiba Jobin Yvon) equipped with a single grating monochromator and photon counting detector TBX-04. An X-ray tube (Seifert, 40 kV, 15 mA) was used for excitation. A powder $Bi_4Ge_3O_{12}$ (BGO) reference scintillator was used for quantitative comparison of the radioluminescence intensity. The spectra were corrected for the spectral dependence of the detection sensitivity.

Photoluminescence (PL) spectra were acquired at room temperature using a compact steady-state spectrofluorometer FluoroMax 4 Plus (Horiba Scientific) equipped with a 150 W ozone-free xenon arc lamp, two Czerny-Turner monochromators, and a photon counting detector R928P PMT. The spectra were corrected for the spectral dependence of the detection and excitation efficiency.

FT-IR spectra were measured using the attenuated total reflectance (ATR) method on a Thermo Scientific Nicolet iS50 FTIR spectrometer.

2.3. Microbiological experiments

The yeast strain *Saccharomyces cerevisiae* haploid type a (DBM 272, ICT Prague, Czech Republic) was used as a model organism. For detailed information on cell preparation, see the Supplementary Information.

To determine dark toxicity of the nanomaterials, *S. cerevisiae* cells were incubated with nanomaterials in physiological saline at 37 °C. Cell viability was determined by counting formed colonies in four replicates; each experiment was repeated three times. For experimental details, see the Supplementary Information.

Based on the dark toxicity results, a nanomaterial concentration of 1 mg/mL was selected for radiosensitisation experiments using X-ray irradiation (see section 2.4.1). The cells (~$10^6$ cells/mL) were incubated with the nanocomposite either for 48 h or irradiated immediately after mixing. Post-irradiation, the formed colonies were counted after 2-day incubation. Full experimental details are provided in the Supplementary Information.

2.4. X-ray irradiation

*2.4.1. X-ray source*

A custom-made research X-ray cabinet, the SCIOX Beam, was used for irradiation. The cabinet is equipped with two wide-angle tungsten X-ray tubes, an oil-based cooling system, a lead-shielded irradiation chamber with dimensions 750 × 750 × 720 mm, a movable shelf controlled by machine software, a ventilation fan in the chamber, a radiation filter holder, a temperature sensor, and an orbital shaker. The parameters of the irradiation in all experiments were 195 kV, 20 mA, 4 mm aluminium filter, shaking velocity 20 RPM, and the distance between the sample and X-ray tube was fixed at 400 mm. The dose rate was determined by Fricke dosimetry (for experimental details, see the Supplementary Information) as (185.0 ± 2.8) Gy/hour. The obtained dose rate reflects the dose absorbed by the Fricke solution in a 15 mL Eppendorf tube, which may slightly differ from nanocomposite / cell suspensions used in experiments.

*2.4.2. Singlet Oxygen Detection*

For detection of singlet oxygen, aminophenyl fluorescein (APF; Thermo Fisher Scientific) and 9,10-anthracenediyl-bis(methylene)dimalonic acid (ABDA; Sigma-Aldrich) probes were used. For APF, a suspension was prepared by mixing 3 mg of nanomaterial in a total volume of 1 mL, which consisted of 42 μL of APF and ethanol (≥99.8%, p.a., PENTA) as the solvent. Ethanol was used to scavenge hydroxyl radicals, as both probes can react with OH radicals. The suspension was placed in Brand macro UV-cuvettes sealed with a stopper and irradiated for 20 and 60 min using the SCIOX Beam irradiator. After irradiation, the photoluminescence spectra were measured in quartz microcuvettes.

For ABDA, a suspension was prepared using 3 mg of nanomaterial in 3 mL of an ethanolic solution of $10^{-3}$ mol/L ABDA. The suspension was irradiated for a total time of 60 min under the same conditions as used with the APF probe. The absorption spectra of the suspension were measured in Brand macro UV-cuvettes.

## 3. Results and discussion

To investigate the various parameters of the nanocomposite system GGAG:Ce@SiO$_2$-RB and the effects of synthesis on its characteristics, the material was prepared at different stages of the synthesis process. Six materials were utilised for evaluation: the core GGAG:Ce ground in a ball mill, the core coated with a silica layer GGAG:Ce@SiO$_2$, material functionalised with photosensitizer Rose Bengal GGAG:Ce@SiO$_2$-RB, materials functionalised with folic acid as a target molecule: GGAG:Ce@SiO$_2$-FA and GGAG:Ce@SiO$_2$-RB-FA, and silica nanoparticles functionalised with photosensitizer Rose Bengal SiO$_2$-RB as a control for radiation-dose enhancement and X-PDT.

The diffraction patterns of all the synthesized materials are shown in **Fig. 1** and are consistent with the cubic garnet structural type (space group Ia$\bar{3}$d). They are all in good agreement with the diffractogram of aluminium gadolinium gallium oxide standard (Gd$_3$Ga$_2$Al$_3$O$_{12}$, record No. 00-046-0448) from the ICDD PDF-2 database. The determined lattice parameters were 12.210±0.003 Å for all the Gd$_3$Ga$_2$Al$_3$O$_{12}$ samples and composites and 12.248±0.005 Å for the Gd$_3$Ga$_{2.5}$Al$_{2.5}$O$_{12}$ ones. These values correspond to the expected lattice parameters (per Vegard's rule) in the Gd$_3$(Ga,Al)$_5$O$_{12}$ solid solution. The average crystallite size determined by a Halder-Wagner method linearization of diffraction peak integral breadths for the Scherrer constant of $K$ = 1.0747 is shown in **Tab. 1**. The recorded diffractograms indicate that neither step of functionalisation affects the core nanoscintillator or its crystallite size.

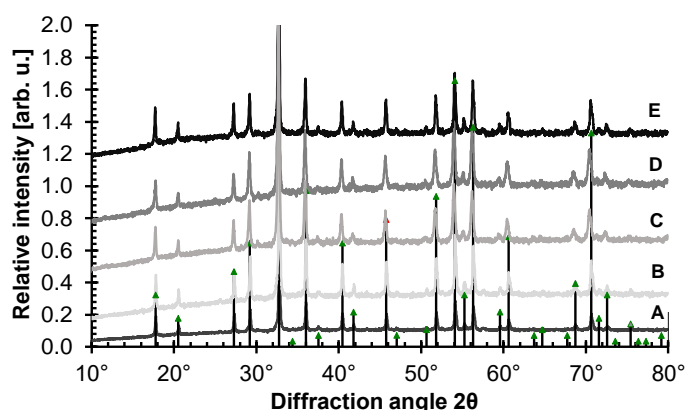

**Fig. 1**. The diffraction patterns of GGAG:Ce (A), GGAG:Ce@SiO$_2$ (B), GGAG:Ce@SiO$_2$-FA (C), GGAG:Ce@SiO$_2$-RB-FA (D), and GGAG:Ce@SiO$_2$-RB (E) compared with standard data of GGAG from ICDD PDF-2 database (record No. 00-046-0448, solid lines). Data are presented with vertical offsets for clarity.

Given the potential medical application of the material, it is crucial to determine its size in a liquid environment. **Tab. 1** and **Fig. SI_2** present the hydrodynamic diameter for all prepared composite materials and SiO$_2$ functionalised with RB. Functionalisation of the material resulted in an increase in the hydrodynamic diameter. The coated nanoparticles GGAG:Ce@SiO$_2$ are approximately 300 nm in diameter, functionalised nanoparticles are approximately 330 nm in diameter, and nanoparticles with folic acid exceed 400 nm in diameter. Similarly, it is also crucial to determine the temporal stability of the nanoparticles in solution; DLS measurements over the span of 90 hours showed a substantial decrease in scattering intensity over time, indicating a limited colloidal stability; these results are presented in the Supplementary Information (see **Fig. SI_3**).

**Tab. 1.** Hydrodynamic diameter and crystallite size of GGAG:Ce, GGAG:Ce@$SiO_2$, GGAG:Ce@$SiO_2$-FA, GGAG:Ce@$SiO_2$-RB-FA, GGAG:Ce@$SiO_2$-RB, and $SiO_2$-RB.

| Material | Hydrodynamic diameter | Crystallite size |
|---|---|---|
| GGAG:Ce | 244 ± 11 nm | 89 ± 21 nm* |
| GGAG:Ce@$SiO_2$ | 302 ± 14 nm | 91 ± 10 nm* |
| GGAG:Ce@$SiO_2$-FA | 409 ± 19 nm | 60 ± 13 nm‡ |
| GGAG:Ce@$SiO_2$-RB-FA | 473 ± 22 nm | 47 ± 10 nm‡ |
| GGAG:Ce@$SiO_2$-RB | 330 ± 15 nm | 74 ± 21 nm* |
| $SiO_2$-RB | 284 ± 13 nm | amorphous |

\* - $Gd_3Ga_2Al_3O_{12}$:Ce, ‡ - $Gd_3Ga_{2.5}Al_{2.5}O_{12}$:Ce

The SEM micrographs of GGAG:Ce@$SiO_2$ (see **Fig. 2**ab) revealed that this nanocomposite consists of both small nanoparticles (~ 50 nm in diameter) and larger agglomerates ranging from ~ 200 nm to a few μm in size. Their equal brightness in the BSE micrographs indicates that all observed objects consist mostly of GGAG with no evidence of separate $SiO_2$ particles. At the same time, no $SiO_2$ layer was observable on the surface of such particles, *i.e.* its thickness was below the SEM resolution. In the GGAG:Ce@$SiO_2$-RB sample (see **Fig. 2**cd), however, there is a discernible layer on the surface of most particles, ca. 50 nm thick, that appears darker in BSE images. Therefore, this thick surface layer probably consists of $SiO_2$ and/or RB molecules.

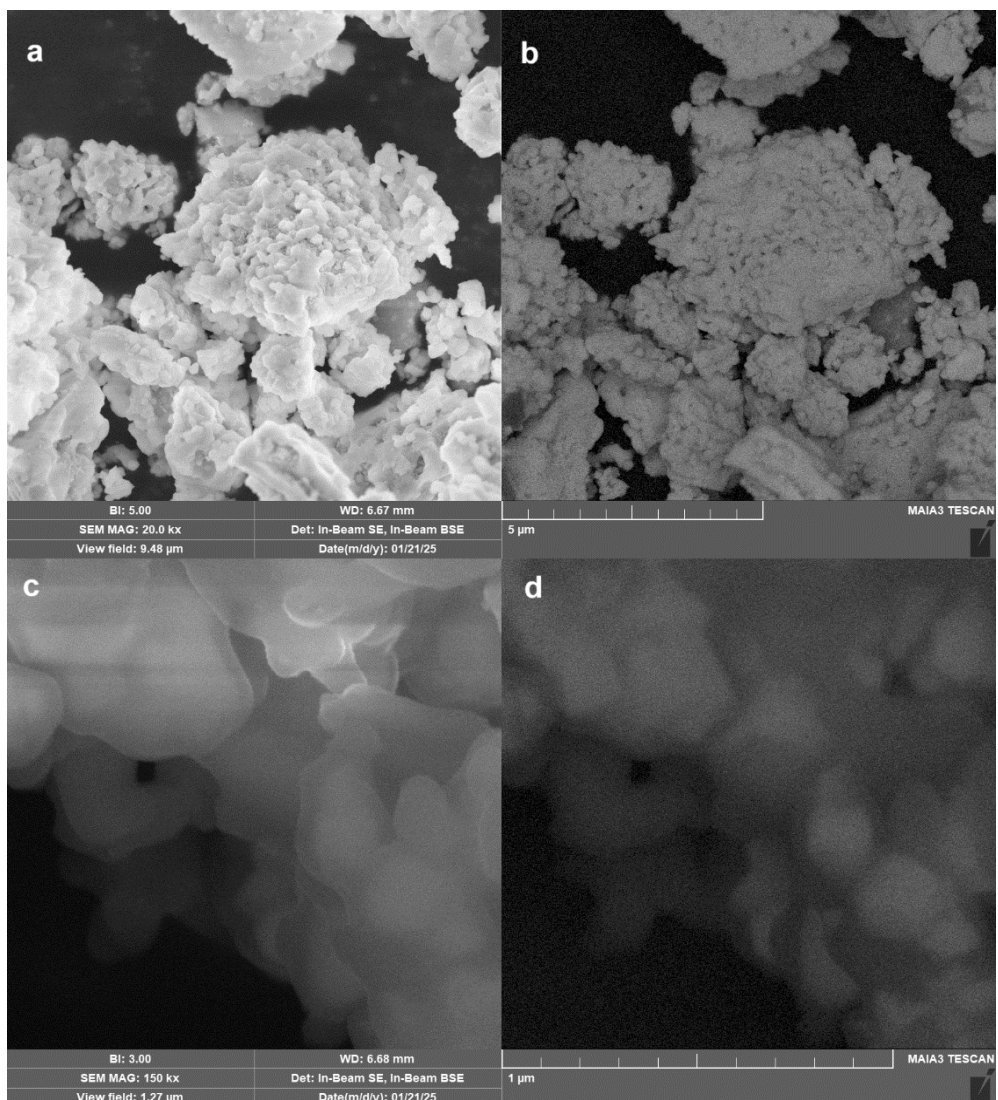

**Fig. 2**. SEM micrographs of GGAG:Ce@SiO$_2$ (a,b) and GGAG:Ce@SiO$_2$-RB (c,d) using either secondary electron detector SE (a,c), or back-scattered electron detector BSE (b,d).

For singlet oxygen production, efficient energy transfer from the scintillation core GGAG:Ce to the photosensitizer Rose Bengal is essential. The Ce$^{3+}$ dopant was selected to ensure that the radioluminescence emission spectrum of the core overlaps with the optical absorption spectrum of Rose Bengal. Radioluminescence spectra allowing quantitative comparison between samples are presented in **Fig. SI_6**. The observed broad emission peak from 500 nm to 700 nm corresponds to the very intense parity-allowed 5d-4f transition of Ce$^{3+}$. The spectrum of GGAG:Ce@SiO$_2$ demonstrates that SiO$_2$ coating causes only a slight decrease in the Ce$^{3+}$ radioluminescence intensity. In GGAG:Ce@SiO$_2$-RB, the luminescence intensity strongly decreased and only a weak emission around 500 nm and 625 nm remained. The folic-acid-functionalized Gd$_3$Ga$_{2.5}$Al$_{2.5}$O$_{12}$:Ce-based composites featured a somewhat lower luminescence intensity compared to the other samples based on Gd$_3$Ga$_2$Al$_3$O$_{12}$:Ce due to the difference in core composition. The GGAG:Ce@SiO$_2$-RB-FA

composite showed only a small decrease of Ce emission around 560 nm compared to GGAG:Ce@SiO$_2$-RB.

Significant overlap of GGAG:Ce radioluminescence and RB absorption spectrum as well as changes in the shape of nanocomposites' radioluminescence spectra are illustrated in **Fig. 3**. The strong decrease in GGAG:Ce@SiO$_2$-RB radioluminescence peak around 570 nm may be attributed to a slightly red-shifted absorption of RB, as absorption and emission shifts have been reported in the literature for RB encapsulation in liposomes (Chang et al., 2008). In both RB-functionalized composites, the intensity ratios compared to a corresponding RB-free material revealed almost identical trough around 565 nm, which we attribute to RB absorption. Lower intensity of this trough in GGAG:Ce@SiO$_2$-RB-FA, i.e. higher luminescence intensity, suggests that biofunctionalization with FA may have either disrupted the energy transfer from garnet core to RB, or decreased the surface concentration of RB.

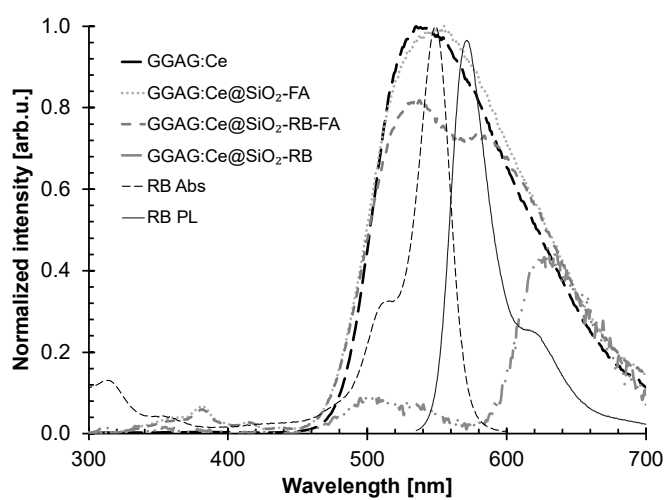

**Fig. 3.** Normalized radioluminescence spectra of GGAG nanocomposites compared to the Rose Bengal (RB) absorption and photoluminescence (PL) spectra in ethanol.

In the photoluminescence (PL) spectra of nanocomposite powders (see **Fig. SI_7**), a typical emission corresponding to Ce$^{3+}$ in garnets was observed when excited by 340 nm or 430 nm (Ce$^{3+}$ 4$f$–5$d$ transitions). Additionally, a weaker emission at ~ 700 nm was also observed that can be attributed to Cr$^{3+}$ emission, which can be present in the samples as unwanted impurity. In the GGAG@SiO$_2$-RB, the garnet emission could not be observed due to its very small intensity, whereas in the GGAG@SiO$_2$-RB-FA a small trough was observed at 560 nm in the cerium emission band similarly to its RL spectra. Despite intense excitation at 550 nm, no emission of RB was observed in any sample, including both GGAG nanocomposites conjugated with RB and the SiO$_2$-RB employed as a control sample. Rose Bengal was unambiguously present in these nanocomposites, as evidenced by their red or pink colour and the FTIR results described below. Therefore, this observation indicates that either the RB luminescence is completely quenched in the solid phase, possibly due to non-radiative losses on a nearby defect, or the process of its conjugation caused a loss of photoluminescence properties. The relatively large thickness of the $SiO_2$/RB layer evidenced in **Fig. 2**cd may have also contributed by enhancing the reabsorption process of the Rose Bengal luminescence. The possibility of non-radiative energy transfer was confirmed by PL decay curves measurement (see **Fig. SI_11**), where a significant acceleration of the PL decays associated with the GGAG:Ce core was observed. Typical radiative decay times of $Ce^{3+}$ in garnets are in the order of tens of nanoseconds, but in RB-functionalized material this component is severely suppressed. The mechanism of energy transfer was already discussed in Popovich et al, (2018) and Procházková et al., (2019).

To further study the functionalization of the surface, the FT-IR spectra of the various nanocomposites were measured. The results confirmed the presence of Rose Bengal in the respective nanocomposite, while the presence of folic acid could not be confirmed with certainty due to a very weak signal. The full spectra are listed in Supplementary Information, **Fig. SI_4** and **Fig. SI_5**.

An appropriate material for radiodynamic therapy should exhibit toxicity only upon irradiation by ionising radiation; thus, for the toxicity study of prepared nanocomposites, *Saccharomyces cerevisiae* was selected as the model organism. The toxicity of the material without irradiation is referred to as dark toxicity. Dark toxicity of GGAG:Ce@$SiO_2$ and GGAG:Ce@$SiO_2$-RB is presented in **Fig. SI_8** and **Fig. 4**. The toxicity immediately following exposure of the material to the cells was compared with the toxicity after 48 h contact time. The 48 h contact time was determined from a time-dependent experiment, in which non-toxic $SiO_2$ nanoparticles were utilised and cells were maintained in contact with nanoparticles. The viability of the cells was observed over 3 days, and based on cell viability, a 48 h contact time was selected. The concentration of $SiO_2$ was chosen according to (Sponza & Nefise, 2020). The concentration of GGAG:Ce@$SiO_2$ and GGAG:Ce@$SiO_2$-RB up to 10 mg per mL (see **Fig. SI_8** and **Fig. 4**) demonstrates negligible toxicity to cells. This provides a suitable foundation for investigating the toxicity after irradiation and toxicity to cancer cells. Dark toxicity of $SiO_2$-RB towards *Saccharomyces cerevisiae* was also found to be negligible (See **Fig. SI_9**).

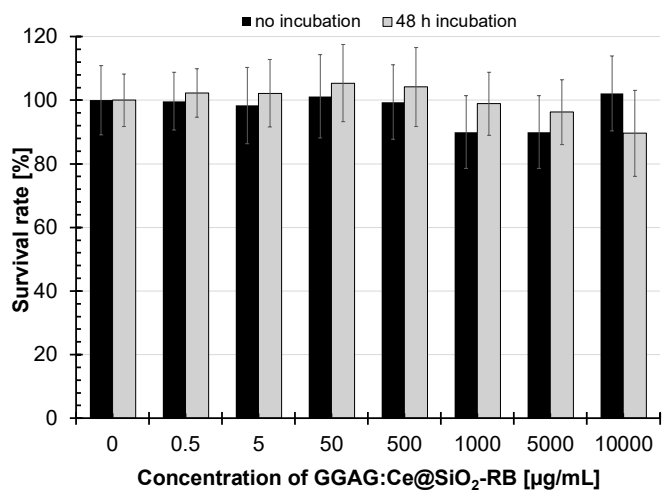

**Fig. 4.** Dark toxicity of GGAG:Ce@SiO$_2$-RB toward *Saccharomyces cerevisiae*.

Following the determination of dark toxicity, the toxicity after X-ray irradiation was investigated. As determined from dark toxicity experiments, the concentration of 1 mg/mL was found to be non-toxic to *S. cerevisiae*; consequently, this concentration was selected for radiosensitisation experiments. The results are presented in **Fig. 5**. The viability of the cells after irradiation with GGAG:Ce@SiO$_2$-RB and without the nanoparticles is comparable, indicating that an insufficient quantity of singlet oxygen was produced to induce cell death. This might result from a negligible nanoparticle uptake into the cells, so an experiment with 48-hour incubation, or contact time, was attempted. Although the survival rate decreased in the presence of nanoparticles compared to nanoparticle-free solutions, the decrease is statistically inconclusive. **Fig. SI_10** presents a microscopic image of the cells after 48 h incubation time. It is observed that nanoparticles are present in close proximity to the cells but are not incorporated into them. From this observation, it can be inferred that there is insufficient singlet oxygen generated near the cells to disrupt the cell wall. The second mechanism of cell damage during X-PDT is radiation-dose enhancement caused by heavy elements, i.e. high atomic number, which constitute nanocomposite core. Due to heavy elements, the X-ray produce more secondary radiation and reactive oxygen species (ROS) leading to cell damage. Since this phenomenon is not extensively observed, it is plausible that the quantity of the material selected was insufficient and a larger concentration of nanoparticles needs to be utilized.

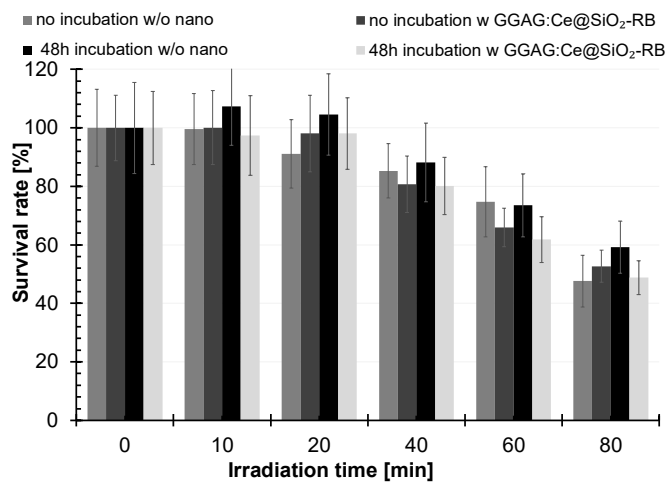

**Fig. 5.** Toxicity of GGAG:Ce@SiO$_2$-RB after X-ray irradiation without incubation and after 48-hour incubation toward *Saccharomyces cerevisiae* compared to a control.

To test singlet oxygen production by the nanocomposite under X-ray irradiation, ABDA and APF chemical probes were used. The reaction of ABDA with singlet oxygen leads to the formation of an endoperoxide, which is indicated by a decrease in the probe's absorbance. Similarly, the reaction of APF with singlet oxygen results in the formation of a fluorescent product with an emission maximum around 520 nm. The results of irradiating the nanocomposites with these probes were always compared to spectra obtained from irradiation of the probes alone.

In the case of ABDA, the results of 40 minutes of irradiation in ethanol with and without the GGAG:Ce@SiO$_2$-RB nanocomposite are shown in **Fig. 6**. Additionally, the nanocomposite without the photosensitizer (GGAG:Ce@SiO$_2$) was also irradiated with the probe for comparison. While the probe alone and the probe with GGAG:Ce@SiO$_2$ showed only a minimal response, a noticeable decrease in absorbance at the 397 nm maximum of the probe was observed during irradiation of the GGAG:Ce@SiO$_2$-RB nanocomposite. However, caution should be exercised when assuming the specificity of the probes for singlet oxygen, especially in the presence of radiolytic products.

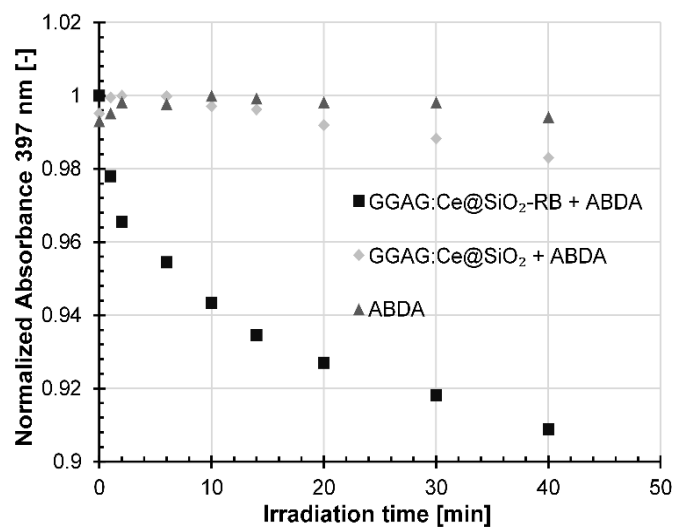

**Fig. 6**. Changes in absorbance at 397 nm of ethanolic solution of GGAG:Ce@SiO$_2$-RB + ABDA, GGAG:Ce@SiO$_2$ + ABDA and ABDA with irradiation time, normalized to the absorbance of a non-irradiated sample.

To address this, the APF probe was used for comparison. The samples were irradiated for 20 and 60 minutes. Additionally, a similar nanocomposite, LuAG:Pr@SiO$_2$-PPIX, previously described by Popovich et al. (2018), was also irradiated with APF under the same conditions for comparison. If singlet oxygen or any other reactive species were interacting with the probe, an increase in APF's emission intensity would be expected with prolonged irradiation. However, as shown in **Fig. 7**, the probe's response to the nanocomposite discussed here did not exceed that observed when irradiated alone. In contrast, a noticeable difference was observed for LuAG:Pr@SiO$_2$-PPIX, clearly proving generation of ROS in this system. The emission bands in the ranges 510-545 nm, 550-650 nm and 615-645 nm in **Fig. 7** belong to APF, Rose Bengal and Protoporphyrin IX, respectively. Contrary to photoluminescence measurements of nanocomposite powders where Rose Bengal emission was not observed, intense RB luminescence was clearly observed in these ethanolic dispersions. It is possible that during the irradiation, some of the bonds linking RB molecules to the nanocomposites were broken by reactive radicals. Based on these results, under the selected irradiation conditions, the generation of singlet oxygen by the GGAG:Ce@SiO$_2$-RB nanocomposite could not be reliably confirmed by either of the probes.

In summary, the findings from the characterisation and biological evaluation suggest that the prepared GGAG:Ce@SiO$_2$-RB does not produce singlet oxygen under X-ray exposure but instead generates other reactive oxygen species (ROS). Further investigation is necessary to evaluate these ROS. The absence of Rose Bengal emission in the photoluminescence spectra of powder samples may be attributed to a self-absorption in the relatively thick SiO$_2$/RB layer or quenching of the luminescence due to aggregation. Consequently, it can be assumed that the inconclusive results in radiosensitisation experiments were due to an insufficient quantity of the nanocomposite, and a higher concentration of nanoparticles is required to observe radiation-dose enhancement in *Saccharomyces cerevisiae*. As previously reported, multicomponent garnet $Lu_3Al_5O_{12}:Pr^{3+}$ (LuAG) encapsulated in silica and functionalized with photosensitizer Protoporphyrin IX (PPIX) demonstrated clear evidence of singlet oxygen generation when tested with the APF probe under X-ray irradiation (Popovich, et al., 2018). $GGAG:Ce@SiO_2$-RB nanocomposite described in this paper shares several characteristics. Both are multicomponent garnets prepared by photo-induced method, encapsulated in silica layer, functionalized with suitable photosensitizer for potential use in X-PDT, and radioluminescence spectra confirm the energy transfer between core and photosensitizer. The primary distinction between these materials lies in photosensitizer emission observed in photoluminescence measurement. Rose Bengal emission was not observed, whereas Protoporphyrin IX emission was observed for $LuAG@SiO_2$-PPIX. Therefore, future studies should focus on: i) optimising the silica coating and Rose Bengal functionalisation process to reduce the thickness of the silica/RB layer to improve energy transfer efficiency, ii) exploring alternative methods for attaching Rose Bengal that may preserve singlet oxygen generation, iii) conducting biological experiments with higher nanoparticle concentration, iv) refining singlet oxygen and ROS detection methods.

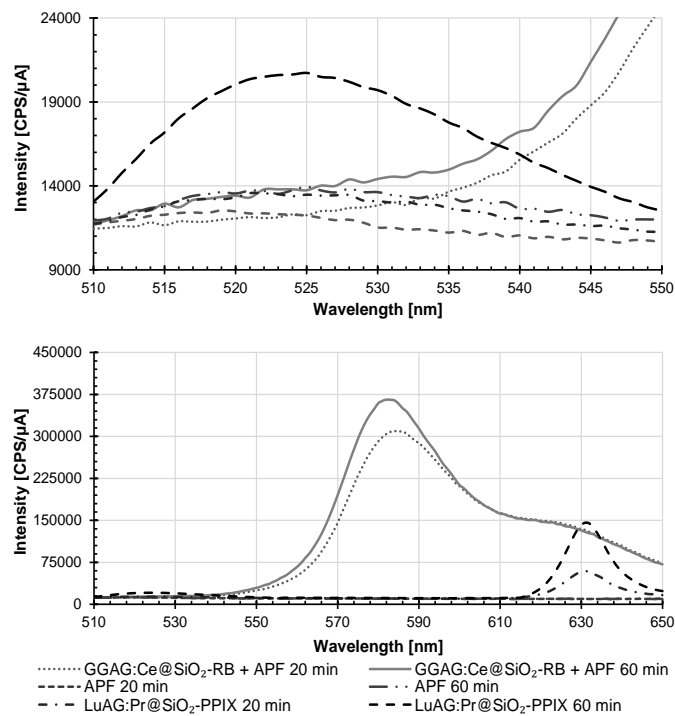

**Fig. 7.** Emission spectra ($\lambda_{ex}$ = 495 nm) of irradiated samples of ethanolic solutions of $GGAG:Ce@SiO_2$-RB + APF, $LuAG:Pr@SiO_2$-PPIX + APF and pure APF (top – APF emission region, bottom – whole spectral range)

## 4. Conclusions

A nanocomposite material based on scintillating multicomponent garnet GGAG doped with Ce was successfully synthesized. The GGAG:Ce core was synthesized via a photo-induced method using aqueous precursor solution and the subsequent calcination of the precipitated product to obtain nanocrystalline particles. Nanoparticles were coated using a sol-gel process by silica layer which was activated by APTES for further functionalization by photosensitizer or target molecule. Conjugation of photosensitizer Rose Bengal or folic acid as target molecule were achieved by EDC/NHS crosslinking. X-ray diffraction confirmed that the cubic garnet structure of the GGAG:Ce core was maintained throughout functionalization steps. Dynamic light scattering showed an increase in hydrodynamic diameter from ~300 nm for GGAG:Ce@$SiO_2$ to ~330 nm after Rose Bengal functionalization. SEM imaging revealed the presence of a ~50 nm thick surface layer on GGAG:Ce@$SiO_2$-RB particles, likely composed of silica and/or Rose Bengal. Both radioluminescence measurements and photoluminescence decays indicate energy transfer from the GGAG:Ce core to Rose Bengal, as evidenced by decreased $Ce^{3+}$ emission and spectral changes. $Ce^{3+}$ core emission was also observed in PL spectra following corresponding excitation. Nevertheless, energy transfer between the core and photosensitiser was not confirmed in PL spectra of powders, as no emission of Rose Bengal was observed under excitation of either the core or Rose Bengal directly. The nanocomposite showed minimal dark toxicity to *Saccharomyces cerevisiae* up to 10 mg/mL concentration. X-ray irradiation experiments did not demonstrate significant singlet oxygen production or radiosensitisation effects compared to controls. Finally, the singlet oxygen generation was examined using ABDA and APF chemical probes as a singlet oxygen indicator. Both probes were inconclusive in detecting singlet oxygen generation under the tested conditions, which were specifically tailored to supress unwanted interference of OH radicals generated in the radiolysis. However, the GGAG:Ce@$SiO_2$-RB composite in ethanol caused a significant decrease in absorbance of the ABDA probe when compared to GGAG:Ce. This observation rules out dose enhancement effect as the cause of this decrease and it is probable that another reactive oxygen species that could not be scavenged by ethanol was generated. While the nanocomposite design shows potential based on its optical properties and low toxicity, further optimisation is needed to achieve effective X-PDT performance. Future work should focus on improving energy transfer efficiency, increasing cellular uptake, and enhancing singlet oxygen generation under X-ray irradiation. Alternative photosensitizers or surface modifications could also be explored to boost the therapeutic efficacy of this nanocomposite system.

**CRediT authorship contribution statement**

**Iveta Terezie Hošnová:** Investigation, Methodology, Writing – original draft, Visualization**, Kristýna Havlinová:** Investigation, Writing – original draft, Visualization**, Jan Bárta:** Investigation, Writing – review & editing, Visualization**, Karolína Mocová:** Investigation**, Xenie Lytvynenko:** Investigation**, Lenka Prouzová Procházková:** Investigation, **Vojtěch Kazda:** Investigation**, František Hájek:** Investigation**, Viliam Múčka:** Supervision**, Václav Čuba:** Funding acquisition, Methodology, Writing – review & editing

**Declaration of competing interest**

The authors declare no conflict of interest. The funders had no role in the design of the study; in the collection, analyses, or interpretation of data; in the writing of the manuscript, or in the decision to publish the results.

**Declaration of Generative AI and AI-assisted technologies in the writing process**

During the preparation of this work the author I.T.Hošnová used Paperpal and DeepL Translator in order to check grammar and improve language and readability. After using this tool, the authors reviewed and edited the content as needed and take full responsibility for the content of the publication.

**Acknowledgements**

This work was supported by the Czech Science Foundation under project no. GA23-05615S, by the Czech Technical University in Prague grant No. SGS23/189/OHK4/3T/14, and by Operational Programme Johannes Amos Comenius financed by European Structural and Investment Funds and the Czech Ministry of Education, Youth and Sports, project No. SENDISO - CZ.02.01.01/00/22_008/0004596.

**Data availability**

Data will be made available on request.